\documentclass[
11pt,%
tightenlines,%
twoside,%
onecolumn,%
nofloats,%
nobibnotes,%
nofootinbib,%
superscriptaddress,%
noshowpacs,%
centertags]%
{revtex4}
\usepackage{ulem}
\usepackage[dvipsnames]{xcolor}
\usepackage{tikz}

\definecolor{mygreen}{RGB}{0, 185, 70}

\begin{document}


\title{Towards ML-based diagnostics of focused laser pulse}

\author{\firstname{Y.~R.}~\surname{Rodimkov}}
\email[E-mail: ]{rodimkov@bk.ru}
\affiliation{Department of Mathematical Software and Supercomputing Technologies, Lobachevsky University, 603950 Nizhni Novgorod, Russia}

\author{\firstname{V.~D.}~\surname{Volokitin}}
\email[E-mail: ]{valyav95@mail.ru}
\affiliation{Department of Mathematical Software and Supercomputing Technologies, Lobachevsky University, 603950 Nizhni Novgorod, Russia}
\affiliation{Mathematical Center, Lobachevsky University, 603950 Nizhni Novgorod, Russia}

\author{\firstname{I.~B.}~\surname{Meyerov}}
\email[E-mail: ]{meerov@vmk.unn.ru}
\affiliation{Department of Mathematical Software and Supercomputing Technologies, Lobachevsky University, 603950 Nizhni Novgorod, Russia}
\affiliation{Mathematical Center, Lobachevsky University, 603950 Nizhni Novgorod, Russia}

\author{\firstname{E.~S.}~\surname{Efimenko}}
\email[E-mail: ]{evgeny.efimenko@ipfran.ru}
\affiliation{Institute of Applied Physics of the Russian Academy of Sciences, 603950 Nizhni Novgorod, Russia}



\begin{abstract} 
Currently, machine learning (ML) methods are widely used to process the results of physical experiments. In some cases, due to the limited amount of experimental data, ML-models can be pre-trained on synthetic data simulated based on the analytical theory and then fine-tuned using experimental data. A limitation of this approach is the presence of the latent parameters of the analytical model, which values are difficult or impossible to estimate. Setting these parameters incorrectly may induce a dataset shift even when applied to synthetic data. To overcome this problem, we train the ML-model on a dataset with randomly varied latent parameters of the analythical model to force the ML-model to concentrate on more general patterns that depend weakly on the latent parameters.
We applied this approach to the problem of tight focusing of a laser pulse with the complex structure of the wavefront. We observed good accuracy of reconstructing of the tilt parameters when training and testing the ML-model on datasets generated for different values of the latent parameters. This confirms that the ML-model was able to select relevant information without over-fitting for specific features inherent in certain values of the latent parameters. We believe that this approach will enrich possible applications of ML-methods to an experimental diagnostics of laser pulses.
\end{abstract}



\maketitle

\section{Introduction}

With the progress in optics technologies in the last decades, laser facilities can now achieve unprecedented intensities. Femtosecond laser pulses focused to spot size of the order of a few micrometers can now routinely achieve intensities above $10^{22}$~W/cm$^{2}$~\citep{Yanovsky2008,Pirozhkov2017,Tiwari2019}, with a record intensity even exceeding $10^{23}$~W/cm$^{2}$~\cite{Yoon2021}. These ultra-relativistic laser pulses allow for multiple applications such as particle acceleration in plasmas~\cite{Higginson2018,Gonsalves2019} and production of $\gamma$-rays~\cite{Sarri2014}, or exploring quanto-electrodynamic effects in laser-plasma interaction in the high-field regime~\cite{DiPiazza2012}.

Such intensities are reached by the OPCPA technique, which successively stretches and compresses the pulse for further amplification~\citep{Strickland85}. Any imperfection in this process in the laser chain can thus introduce spatio-temporal couplings (STC), i.e. correlations between the longitudinal and transverse intensity profiles, which ultimately reduces the intensity at focus~\citep{Kim2013, Bourassin2013}. STC tends to become stronger for more powerful laser systems -- corresponding to an increasing waste of the pumping energy, which becomes greatly harmful for applications.

Characterization of the laser pulse is thus crucial both for optimizing the laser profile in experiments and for its accurate numerical description without overestimation of the peak intensity -- leading to better agreement with the experimental results and thus ultimately to a deeper understanding of the underlying physics.

Spatio-temporal characterization of the ultra-intense laser pulses is however not straightforward. Direct detection is precluded due to the high frequency of the pulse, and a combination of spectroscopy and interferometry has to be used. 3D characterization is thus a trade-off between cost, time of acquisition and resolution of the data. Recently, experimental techniques leading to a complete spatio-temporal characterization of ultra-intense laser pulses have been developed, such as TERMITES~\cite{Pariente2016} or INSIGHT~\cite{Borot2018}, which have been applied to top-class PW-laser facilities~\cite{Jeandet2019}. Such measurements are however obtained thought extensive manipulations, and effort are also pursued for simplifying this process through the combination of experimental measurements and machine learning. First attempts in this sense were realized in the 90's, in which neural network~\cite{Krumbugel1996} and genetic algorithm~\cite{Nicholson1999} were used with frequency-resolved optical gating (FROG) for reconstructing the pulse phase. More recently, a neural network trained on simulated data retrieved the pulse phase even in the presence of high noise, and lowered the required knowledge about the relation between the pulse and its measurement~\cite{Zahavy2018}. Other works have also proposed to reconstruct the pulse phase with neural network from dispersion scan traces~\cite{Kleinert2019}, 2D intensity patterns~\cite{Ziv2020} or with a multimode fiber~\cite{Xiong2020}.

Training data for machine learning (ML) models can be collected experimentally or through simulation. Quite often the amount of experimental data is not sufficient for training the ML-model from scratch. In this case the training data can be numerically simulated on the basis of some analytical theory, and then fine-tuned on a real experimental data. This approach, although seems very natural, requires good correspondence between the analytical theory and experiment. The later can become very problematic, because the analytical model may have hidden or latent parameters which affect the results, but can not be directly measured or estimated in the experiment. The number of such latent parameters for ML-models can be large. Setting parameters incorrectly during data collection can cause a ML-model to break when applied to real data due to dataset shift.


In this article, we demonstrate an approach to reconstructing the physical parameters of a tightly focused laser pulse based on the energy flux distribution in the case of the latent parameters uncertainty. In order to make the laser pulse model closer to experimental conditions we impose a spectral-dependent tilt on its wavefront, and the properties of this tilt act as the latent parameters of the model. The resulting analytical model is rather simplified representation of the real experimental conditions, nevertheless it can help to find better approaches for the problem. This parameterized model with randomly varied parameters was used for synthetic data generation.  We use this data to train the ML-model to solve the inverse problem of reconstructing pulse parameters. In order to study the influence of the choice of the  latent parameters on accuracy of the ML-model we study the generalization ability of this model by training and testing on different subsets of the latent parameters values. By doing this we demonstrate tolerance of the ML-model to wrong choice of the latent parameters during training, that can be useful for the application of ML-methods to real experimental data. 

The article is structured as follows. In Section~\ref{sec:generative} the analytical model used for data generation is described. In Section~\ref{sec:results} we present the methodology of our work, methods and metrics used in this paper, the solution of the inverse problem and the analysis of the generalization ability of the ML-model.  In Section~\ref{sec:discussion} the results are discussed and some considerations related to further development of the approach are given. Finally, in Section~\ref{sec:conclusions} the conclusions are formulated. 


\section{Generative model}
\label{sec:generative}

The data was generated by simulation of the propagation of a tightly focused laser pulse initialized in the far field zone to the focal plane. The modeling was performed by means of the Hi-chi module~\citep{Panova2020}, which uses a spectral method for solving Maxwell's equations. The cumulative energy flux was calculated in the focal plane to be used for the ML-model training.

For simplicity the pulse propagates along the $x$-axis and the pulse wavefront is chosen to be spherical, so any field component $u$ in the far field zone can be set as 
\begin{equation}
u(x,y,z) = u_l(r(x,y,z))u_t(\alpha(x,y,z)),
\label{eq:general}
\end{equation}
where $r = |\mathbf{r}| = |\mathbf{R} - \mathbf{R}_0|$ is the distance from the center of the spherical wavefront $\mathbf{R}_0$ to the point $\mathbf{R}$ and $\alpha = \arcsin \sqrt{y^2+z^2}/r$ is the angle between the $x$-axis and vector $\mathbf{r}$. 
The electromagnetic fields are defined as $\mathbf{E}(\mathbf{R})=u(\mathbf{R})\mathbf{s}_0$ and $\mathbf{B}(\mathbf{R})=u(\mathbf{R})\mathbf{s}_1$ with the normalized vectors $\mathbf{s}_0=\mathbf{e_y}\times \mathbf{e_y}\times \mathbf{r}$ and $\mathbf{s}_1=\mathbf{e_y}\times \mathbf{r}$.

The longitudinal $u_l$ and transverse $u_t$ profiles are defined as follows
\begin{equation}
    u_{\rm{l}}(x) = \sin\left(\frac{2\pi x}{\lambda_0}+\varphi\right) \exp\left(-\frac{x^2}{2L^2}\right) \Pi\left(x,-4L,4L\right),
    \label{eq:longitudinal}
\end{equation}
\begin{equation}
    u_{\rm{t}}(\alpha) = \Pi\left(\alpha,0,\theta-\frac{\varepsilon}{2}\right)+\cos^2\left(\frac{\pi(\alpha-\theta+\varepsilon/2)}{2\varepsilon}\right)\Pi\left(\alpha,\theta-\frac{\varepsilon}{2},\theta+\frac{\varepsilon}{2}\right),
\end{equation}
where $\lambda_0$ is the central wavelength, $L$ is the length of the pulse and $\Pi$ is the rectangular function, with $\Pi(x,x_{min},x_{max}) = 1$ if $x$ is in $[x_{min},x_{max}]$ and 0 otherwise. In this case $u_{\rm{t}}$ corresponds to a transverse flattop profile with no intensity outside of the opening angle $\theta$ with the edge smoothing angle $\varepsilon$. The opening angle $\theta$ is connected to the F-number of the focusing optics by $\theta = \arctan(1/2\mathrm{F})$. An additional phase $\varphi$ is used to model the imperfections of the wavefront that occur in experimental conditions. 

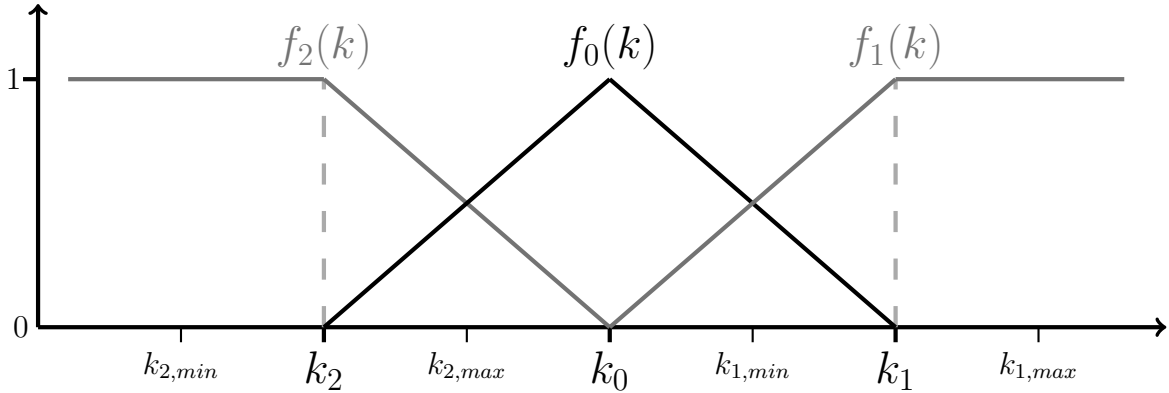
\begin{figure}[t!]
\newcommand\xaxis{15}
\newcommand\yaxis{4.3}

\newcommand\zero{0}

\newcommand\ylim{3.3}
\newcommand\step{0.2}

\newcommand\wrstep{3.8}

\definecolor{grey1}{RGB}{115, 115, 115}
\definecolor{grey2}{RGB}{170, 170, 170}

\begin{tikzpicture}
\draw[ultra thick, ->, black] (\zero, \zero) -- (\xaxis, \zero);
\draw[ultra thick, ->, black] (\zero, \zero) -- (\zero, \yaxis);

\draw[ultra thick, -, black] (\zero, \ylim) -- (-\step, \ylim);

\draw[ultra thick, -, black] (\wrstep, \zero) -- (\wrstep, -\step);
\draw[ultra thick, -, black] (\wrstep*2, \zero) -- (\wrstep*2, -\step);
\draw[ultra thick, -, black] (\wrstep*3, \zero) -- (\wrstep*3, -\step);

\draw[thick, -, black] (\wrstep*0.5, 0) -- (\wrstep*0.5, -\step);
\draw[thick, -, black] (\wrstep*1.5, 0) -- (\wrstep*1.5, -\step);
\draw[thick, -, black] (\wrstep*2.5, 0) -- (\wrstep*2.5, -\step);
\draw[thick, -, black] (\wrstep*3.5, 0) -- (\wrstep*3.5, -\step);

\node[below] at (\wrstep, -\step) {\color{black} \LARGE $k_2$};
\node[below] at (\wrstep*2, -\step) {\color{black} \LARGE $k_0$};
\node[below] at (\wrstep*3, -\step) {\color{black} \LARGE $k_1$};

\node[below] at (\wrstep*0.5, -\step) {\color{black} \large $k_{2, min}$};
\node[below] at (\wrstep*1.5, -\step) {\color{black} \large $k_{2, max}$};
\node[below] at (\wrstep*2.5, -\step) {\color{black} \large $k_{1, min}$};
\node[below] at (\wrstep*3.5, -\step) {\color{black} \large $k_{1, max}$};

\draw[ultra thick, dash pattern=on 8pt off 10pt, grey2] (\wrstep, \zero) -- (\wrstep, \ylim);
\draw[ultra thick, dash pattern=on 8pt off 10pt, grey2] (\wrstep*3, \zero) -- (\wrstep*3, \ylim);

\draw[ultra thick, -, grey1] (\step*2, \ylim) -- (\wrstep, \ylim);
\draw[ultra thick, -, grey1] (\wrstep, \ylim) -- (\wrstep*2, \zero);

\draw[ultra thick, -, black] (\wrstep, \zero) -- (\wrstep*2, \ylim);
\draw[ultra thick, -, black] (\wrstep*2, \ylim) -- (\wrstep*3, \zero);

\draw[ultra thick, -, grey1] (\wrstep*2, \zero) -- (\wrstep*3, \ylim);
\draw[ultra thick, -, grey1] (\wrstep*3, \ylim) -- (\wrstep*3.8, \ylim);

\node[left] at (-\step*0.5,\ylim) {\color{black} \large $1$};
\node[left] at (\zero, \zero) {\color{black} \large $0$};

\node[above] at (\wrstep,\ylim) {\color{grey1} \LARGE $f_2(k)$};
\node[above] at (\wrstep*2,\ylim) {\color{black} \LARGE $f_0(k)$};
\node[above] at (\wrstep*3,\ylim) {\color{grey1} \LARGE $f_1(k)$};

\end{tikzpicture}
\caption{\label{fig:mask_func} Mask functions $f_i(k)$ used to create three spectral components with different tilt properties.}
\end{figure}

The proposed pulse definition does not account for complex structure of wavefront inherent to real experimental conditions. In order to model complex wavefront structure in a simplified manner we introduce three spectral components by applying mask functions $f_i(k)$ (see Fig.~\ref{fig:mask_func}) to the spectra of the longitudinal profile $u_{\rm{l},i}(k)=u_{\rm{l}}(k)f_i(k)$, where $u_{\rm{l}}(k) = \int_{-\infty}^{\infty}u_{\rm{l}}(x)e^{-ik x}dx$ is the spatial Fourier spectra of the longitudinal profile $u_{\rm{l}}(x)$ and $k$ is the wavenumber. Three spectral components are separated by the two wavenumbers $k_1$ and $k_2$ such as $k_1>k_0>k_2$ with $k_i=2\pi c/\lambda_i$, where $c$ is the speed of light. These two boundary wavenumbers $k_{1,2}$ are varied in the range $[k_{i,\rm{min}},k_{i,\rm{max}}$] to parameterize the laser pulse definition. For convenience, later in the text we operate with corresponding wavelengths $\lambda_{1,2}$ as boundary wavelengths and treat them as the latent parameters of the model. 

For each of these three spectral component we add a linear tilt by adding the additional phase $\varphi_i$ in the longitudinal profile $u_i(x)$ in the Eq.~\ref{eq:longitudinal}
\begin{equation}
\varphi_i=a_i y^{\prime}_i,~~ y^{\prime}_i = y\cos\theta_i-z\sin\theta_i,
\label{eq:phase}
\end{equation}
where $a_i$ is a constant characterizing the amplitude of the tilt and $\theta_i$ is an angle characterizing the direction of the tilt in the transverse plane. Further in this paper $a_i$ and $\theta_i$ are referred to as the tilt amplitude and the tilt angle, respectively.

After applying the mask function we calculate longitudinal profile $u_{\rm{l},i}$ for each spectral component by performing inverse Fourier transform $u_{\rm{l},i}(x) = 1/2\pi\int_{-\infty}^{\infty}u_{\rm{l}}(x)e^{ik x}dk$ and calculate the corresponding spectral component initial distribution $u_i$ according to Eq.~\ref{eq:general}
\begin{equation}
    u_i(x,y,z) = u_{\rm{l},i}(r)~u_{\rm{t}} (\alpha).
\end{equation}
 
After initialization in the far field zone each component is propagated to the focal plane using the spectral solver. Using the linearity of Maxwell's equations, we can then get the total field at the focal position as the sum of fields of all spectral components as
\begin{equation}
    u_{\rm{tot}}(\mathbf{R}) =\sum_i  A_i u_i(\mathbf{R}),
\end{equation}
with $A_i$ defining the relative spectral component amplitude compared with the central component corresponding to the mask function $f_0$ (i.e. $A_0=1$). These relative amplitude parameters $A_{1,2}$ act as another pair of latent parameters of our laser pulse model.

Finally, we calculate the cumulative energy flux in the focal plane
\begin{equation}
    I(y,z) =\frac{1}{4\pi}\int_{-\infty}^{\infty}\mathbf{E}_{\mathrm{tot}}(x,y,z)\times \mathbf{B}_{\mathrm{tot}}(x,y,z)\cdot \mathbf{e_x} dx.
\end{equation}

Summarizing, in our simplified model the energy flux depends on following parameters: the boundary wavelengths $\lambda_{1,2}$, the relative amplitudes $A_{1,2}$, the tilt amplitudes $a_i$ and the tilt angles $\theta_i$ with $i\in [0,1,2]$. The last six parameters $a_i$ and $\theta_i$ were predicted by the ML-model, while $\lambda_{1,2}$ and $A_{1,2}$ acted as the latent parameters of the pulse model. Data was generated using central wavelength $\lambda_0 = 800~$nm, F-number of the focusing optics F=1.5, the pulse is initialized at $65\lambda_0$ from the spherical wavefront center. Other parameters were set in the following ranges: $\lambda_1 \in [600\mathrm{nm}, 750\mathrm{nm}]$, $\lambda_2 \in [850\mathrm{nm}, 1000\mathrm{nm}]$, $A_{1,2} \in [0.5,2]$, $a_{0,1,2} \in [0,0.6]$ and $\theta_{0,1,2} \in [0,2\pi]$.

\section{Results}
\label{sec:results}

\subsection{Methodology}
The problem under consideration is the inverse problem of predicting of six parameters (the tilt angle and the tilt amplitude for three spectral components) based on the cumulative energy flux in the focal plane. The analytical model which describes the experiment has latent parameters, which can not be directly reconstructed from the experiment. In our rather simplified model such parameters are the boundary wavelengths $\lambda_{1,2}$ and the relative amplitudes of the spectral component with the lowest and highest wavelengths, respectively, $A_{1,2}$. Determination of these parameters in an experiment, in which the ML-model will be used after training on synthetic data, is a difficult task. Thus, methodologically, the work is divided into two stages. At the first stage, the inverse problem of reconstructing the tilt parameters $a_i$ and $\theta_i$ is solved. In order to do this a suitable architecture of the ML-model is chosen, and the best way for reconstructing the parameters is searched for.  At the second stage, after the inverse problem is solved, the generalization ability of the proposed ML-model is analyzed. To do this, the ML-model is trained on a subset of latent parameters values, and then tested on the entire available set of latent parameter values, i.e. the test set includes values of latent parameters that were not present in the training set. Such experiments will help to answer the main question of interest of this paper, whether the trained ML-model is able to generalize the extracted information to unknown values of latent parameters or not. The data and scripts required to reproduce the numerical results may be downloaded from https://github.com/hi-chi/Machine-Learning (the relevant examples are located in the "Focus ML" folder).

\subsection{Methods and metrics} \label{sec:methods}
The cumulative energy flux in the focal plane can be treated as a single-channel image, see Fig.~\ref{fig:examples}. Neural networks are one of the most versatile methods for analyzing data with the same type of continuous features. For image analysis, the best results at the moment are shown by convolutional neural networks~\citep{LeCun1998, Krizhevsky2012} and transformers~\citep{Vaswani2017, Dosovitskiy2020}. It is known that transformers are difficult to train, they are very demanding on the amount of data, and also require pre-training. At the moment, in computer vision problems, there are not a large number of successful applications of this method without model pre-training~\citep{Caron2021, Touvron2021}. Therefore, convolutional neural networks were chosen for the ML-based analysis of the flux.

After several experiments a convolutional neural net architecture was chosen that consists of two consecutive blocks. Each block included two convolutional layers and a max pooling layer, with the number of convolutions 16 and 32 in each block, respectively. These two blocks were followed by four fully connected layers with the number of neurons 1024, 512, 128, 32, and finally there was a layer that predicts 6 tilt parameters. ReLU was used as an activation function on each convolutional and fully connected layer, except for the last one. For training, the root-mean-square error and the Adam optimizer with default parameters in the Keras framework were used. The model was trained for 30 epochs.

To assess the quality of ML-models in our study we computed the Mean Absolute Percentage Error (MAPE) and the coefficient of determination ($R^2$):
\begin{equation}
    MAPE = \frac{100}{n}  \sum\limits_{i=1}^n\frac{|\widehat{y_i} - y_i|}{\max(\widehat{y})},
\end{equation}    
\begin{equation}
    R^2=1-\sum\limits_{i=1}^n\frac{(\widehat{y_i}-y_i)^2}{(\overline{y}-y_i)^2},
\end{equation}
where $\widehat{y_i}$ and $y_i$ are the true and the predicted value of the $i$-th object, respectively, and $\overline{y}$ is the value of the predicted parameter averaged over the training set.

For the reconstructed parameters a linear transformation to the range from -1 to 1 was used. Each energy flux profile was normalized in the range from zero to one. In all experiments Keras and Tensorflow frameworks were used to train neural networks.

\subsection{The solution of the inverse problem}
\label{sec:inverse}
Our goal was to reconstruct the tilt amplitude $a_i$ and the tilt angles $\theta_i$ for three spectral components of a laser pulse based on the profile of the cumulative energy flux in the focal plane. The study of the data and experiments showed that in order to achieve good accuracy the reconstructed parameters should be carefully chosen. First, the tilt angles $\theta_i$ have an obvious periodicity with a period of $2\pi$, and therefore require a special treatment. There are different approaches to this problem, for example, using a special loss function that takes into account the periodicity of angles, or reconstructing trigonometric functions of angles instead of angles themselves. Second, when the tilt amplitude $a_i$ is small, the dependence of the flux on the corresponding tilt angle $\theta_i$ becomes weak, i.e. specific angle values are indistinguishable for the ML-model, which leads to a strong increase in the angles-related error. This is caused by the fact that the tilt angle and amplitude are not used separately in computation of the additional phase $\varphi_i$ (see Eq.~\ref{eq:phase}), but as the product of the tilt amplitude and the trigonometric functions of the tilt angle. Experiments showed that instead of reconstructing six original parameters $a_i$ and $\theta_i$ it makes sense to reconstruct six derived parameters that are the combination of the original ones $p_i = a_i \sin \theta_i$, $q_i = a_i \cos \theta_i$. This transformation can be interpreted as a transition from polar coordinates to Cartesian coordinates. It should be noted that the proposed transformation is one-to-one, and the original parameters $a_i$ and $\theta_i$ can be reconstructed from derived parameters, but the accuracy of angle reconstructing for the close to zero tilt amplitude can still be poor.

The ML-model architecture and training parameters are described in Section 3.2. 
For the ML-model training an experimental dataset of 50000 samples was numerically generated. The experimental results showed that the parameters $p_i$, $q_i$ are reconstructed with an accuracy close to ideal (see Table \ref{table:accuracy}). It should be noted that the parameters $p_1$ and $q_1$ are reconstructed worse than other parameters, although the accuracy is still high. This effect is systematic and was observed in all experiments. Apparently, this result is due to the fact that these parameters correspond to the spectral component with the shortest wavelength, information about which is worse extracted from the cumulative energy flux. The accuracy of reconstructing of the parameters related to the central spectral component $p_0$ and $q_0$ is the best, which can be explained by the fact that both its amplitude and the central wavelength are fixed. Thus, the use of a convolutional network makes it possible to solve the inverse problem with a very high accuracy.

\begin{table}[h!]
    \caption{Average values and standard deviations of the coefficient of determination (R$^2$) and the mean error (MAPE) for reconstructing of derived parameters $p_i$ and $q_i$ by trained convolutional network. Metrics were averaged over 10 runs with different splits into training and test samples in a 90/10 ratio.}
    \begin{center}
    \begin{tabular}{c|c|c|c|c|c|c} 
    \hline
        Metrics & $p_0$ & $p_1$ & $p_2$ & $q_0$ & $q_1$ & $q_2$ \\ [0.1ex] 
    \hline\hline
    $R^2$ & $0.991\pm0.001$  & $0.948\pm0.006$  & $0.981\pm0.003$ & $0.991\pm0.002$ & $0.951\pm0.006$ & $0.982\pm0.003$ \\ 
    \hline
    MAPE, \% & $2.71\pm0.21$ & $5.70\pm0.42$ & $3.77\pm0.32$ & $2.74\pm0.24$ & $5.66\pm0.45$ & $3.78\pm0.33$ \\
    \hline
    \end{tabular}
\end{center}
\label{table:accuracy}
\end{table}

\subsection{Analysis of the generalization ability of the ML-model}
\label{sec:generalization}
\subsubsection{The main idea.}

In the previous section, we demonstrated the high accuracy of solving the inverse problem using the convolutional neural network trained on model data. At the same time, the application of ML-models trained on synthetic data to the real physical experiment can be problematic, because the wrong choice of values of the latent parameters of the analytical model during training may significantly affect the accuracy of the ML-model on experimental data, which is explained by a change in the data distribution. In this section we address the ability of the ML-model to cope with such negative effects using synthetic test dataset with other values of the latent parameters.

\begin{figure}
    \centering
    \begin{minipage}[h]{0.24\linewidth}
        \center{\includegraphics[width=1\linewidth]{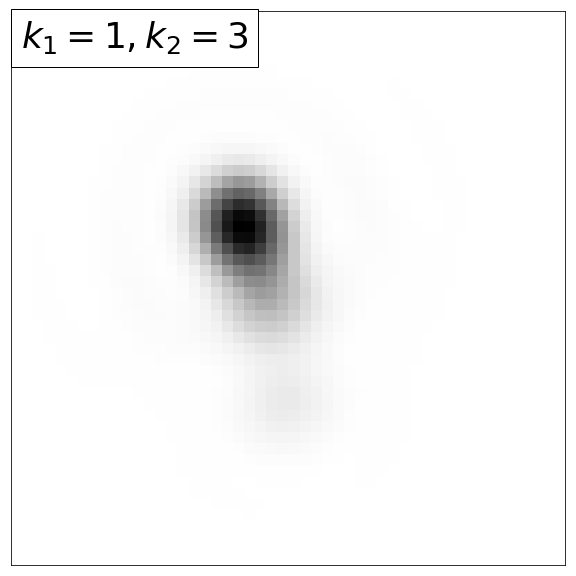} \\ (a)}
    \end{minipage}
    \hfill
    \begin{minipage}[h]{0.24\linewidth}
        \center{\includegraphics[width=1\linewidth]{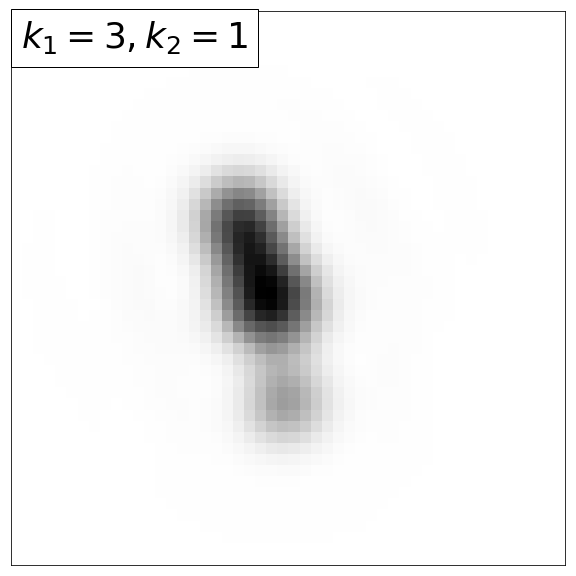} \\ (b)}
    \end{minipage}
    \hfill
    \begin{minipage}[h]{0.24\linewidth}
        \center{\includegraphics[width=1\linewidth]{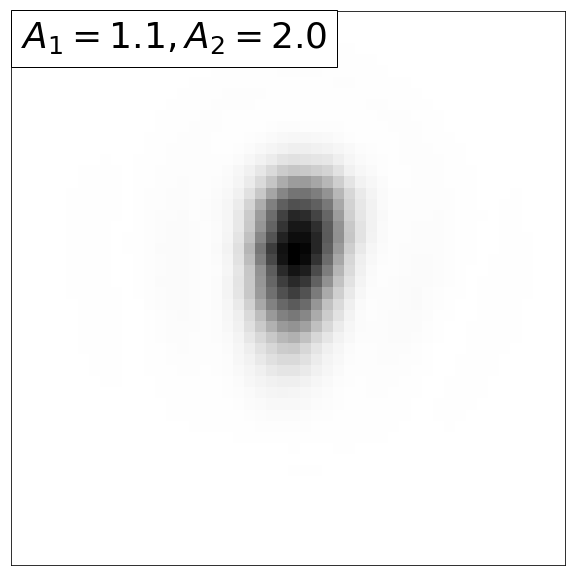} \\ (c)}
    \end{minipage}
    \hfill
    \begin{minipage}[h]{0.24\linewidth}
        \center{\includegraphics[width=1\linewidth]{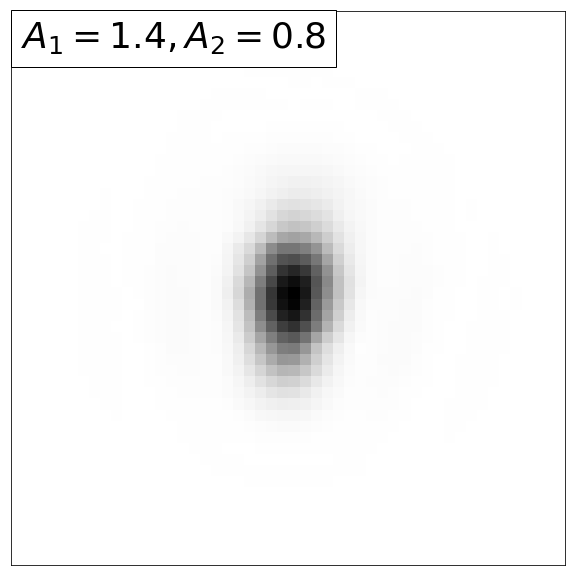} \\ (d)}
    \end{minipage}    

    \caption{Examples of the cumulative energy flux distribution for different values of latent parameters. (a, b) $\lambda_1$ and $\lambda_2$ are different, $A_{1,2}$ are fixed; (c,d) $A_1$ and $A_2$ are different, $\lambda_{1,2}$ are fixed. Parameters $a_i$ and $\theta_i$ are fixed for (a), (b) and (c), (d).}
    \label{fig:examples}
\end{figure}

For a better understanding of the importance of a correct choice of the latent parameters values, it makes sense to visually evaluate the influence of the choice of boundary wavelengths $\lambda_{1,2}$ and relative amplitudes $A_{1,2}$ on the cumulative energy flux profile in the considered analytical model (see Fig.~\ref{fig:examples}). Local changes in the energy distribution can be noticeable, for example the distribution at the bottom in Fig. \ref{fig:examples} (a) is different from the case with the same parameters except $\lambda_{1,2}$ in Fig. \ref{fig:examples} (b). Similar differences can be seen if compare Fig. \ref{fig:examples} (c) and (d) where all parameters are the same except relative amplitudes $A_{1,2}$. These differences show the data variability, but what is more important, is that the general shape of the energy flux distribution in the profile is preserved, which indicates that a significant part of the relevant information is contained in the tilt angles and amplitudes. 

This fact demonstrates the possibility of information generalization for different values of latent parameters, although the good result is not guaranteed. It is known that ML-models can overemphasize irrelevant information. For example, in adversarial attacks, when a change in the data is selected in a special way to change the prediction of the network~\citep{Goodfellow2014, Yuan2019}. There are also examples when irrelevant features were changed in the data, which led to a dataset shift problem and a decrease in the accuracy of the model~\citep{Wang2018, Wilson2020}. In this regard, it seems important to evaluate the influence of the choice of latent parameters on the accuracy of the previously considered ML-model, as well as the ability of the ML-model to extract relevant information even in the case of previously unseen values of the latent parameters.

Based on this, the methodology for further experiments is following: the previously considered convolutional network is trained from scratch on a subset of the data with a reduced set of latent parameter values, and then it is tested on a test dataset that includes all values of latent parameters, including those not present in the training dataset. By doing this we try to quantify the effect of wrong choice of latent parameters values during training on the final accuracy of the ML-model. The experiments  are divided into two parts. In the first part, the boundary wavelengths $\lambda_{1,2}$ act as the studied latent parameters, and in the second part, the same is done for the relative amplitudes $A_{1,2}$.

\subsubsection{Generalization for different wavelengths.}

\begin{figure}
  \centering
  \includegraphics[width=0.75\textwidth]{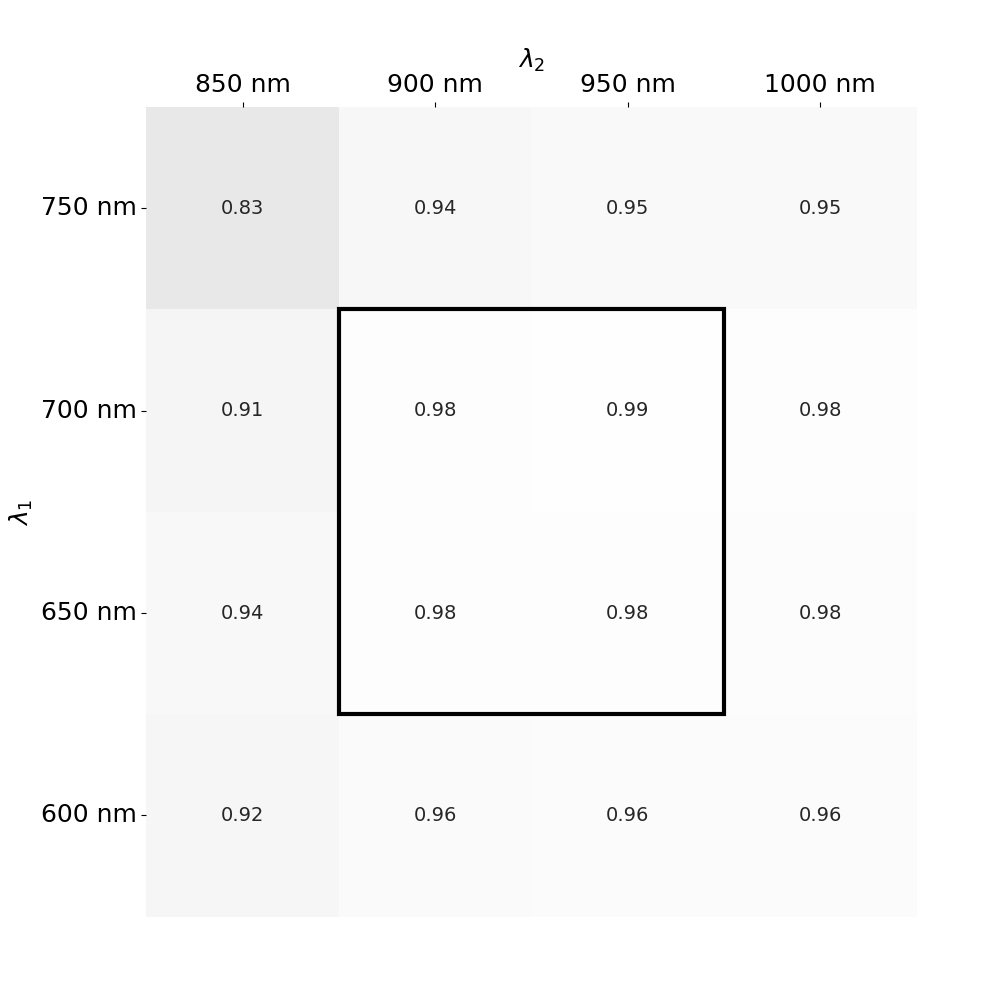}
  
  \caption{The dependence of the coefficient of determination $R^2$ of all parameters on the boundary wavelengths $\lambda_{1,2}$. The coefficient was averaged over 10 runs with random initialization of neural network weights. The black square represents the training data subset. The coefficient of determination is shown by gradation of grey: the lower the coefficient, the darker the color.}
  \label{fig:wavelegths}
  \end{figure}

In the first experiment, the ML-model was trained on a newly generated dataset with 50000 samples with the values of the boundary wavelengths $\lambda_1 \in [650 ~\mathrm{nm}, 700~\mathrm{nm}]$ and $\lambda_2 \in [900 ~\mathrm{nm}, 950~\mathrm{nm}]$. The ML-model was tested on a dataset with 50000 samples from Section 3.3
with $\lambda_1 \in [600 ~\mathrm{nm}, 650 ~\mathrm{nm}, 700 ~\mathrm{nm}, 750~\mathrm{nm}]$ and $\lambda_2 \in [850 ~\mathrm{nm}, 900 ~\mathrm{nm}, 950~\mathrm{nm}, 1000~\mathrm{nm}]$. All other parameters were randomly changed. The model architecture and training parameters are the same as described in Section 3.2.
The results of the experiments are shown in Fig.~\ref{fig:wavelegths}. 

It can be seen that the accuracy of determining the parameters on data with the boundary wavelengths that are not included in the training set slightly decreases. The ML-model shows the worst result when the values of the boundary wavelengths become close to each other. In this case, the difference between the spectral components from different spectral ranges decreases, which complicates the reconstructing of parameter values.

\subsubsection{Generalization for different amplitudes.}

\begin{figure}
    \centering
    \begin{minipage}[h]{0.49\linewidth}
        \center{\includegraphics[width=1\linewidth]{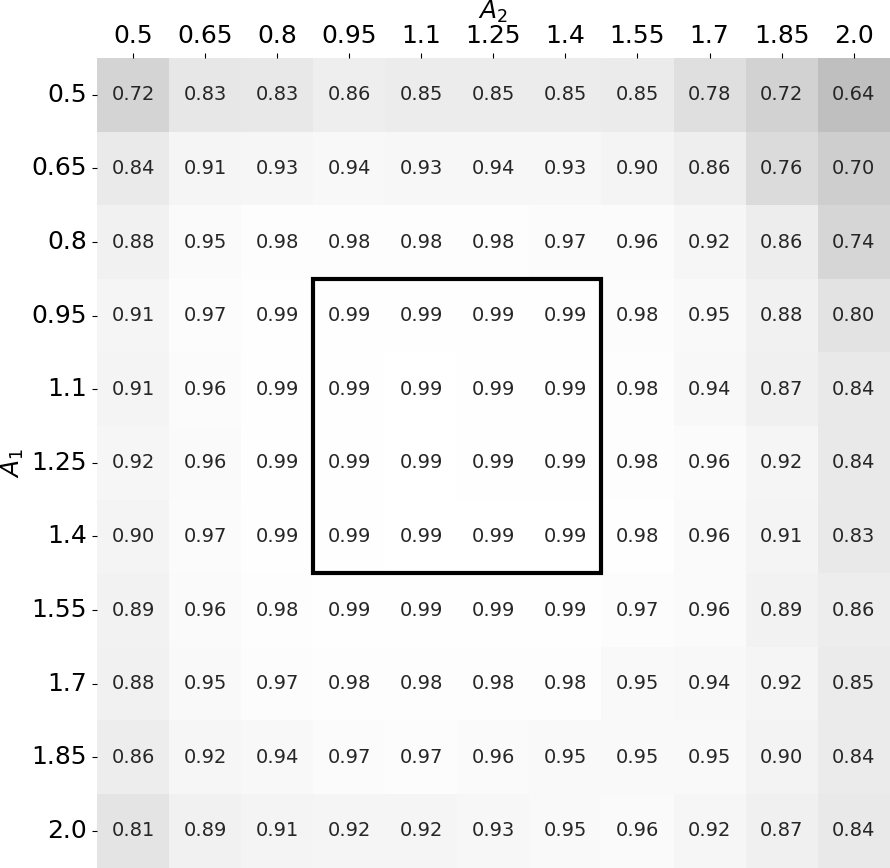} \\ (a)}
    \end{minipage}
    \hfill
    \begin{minipage}[h]{0.49\linewidth}
        \center{\includegraphics[width=1\linewidth]{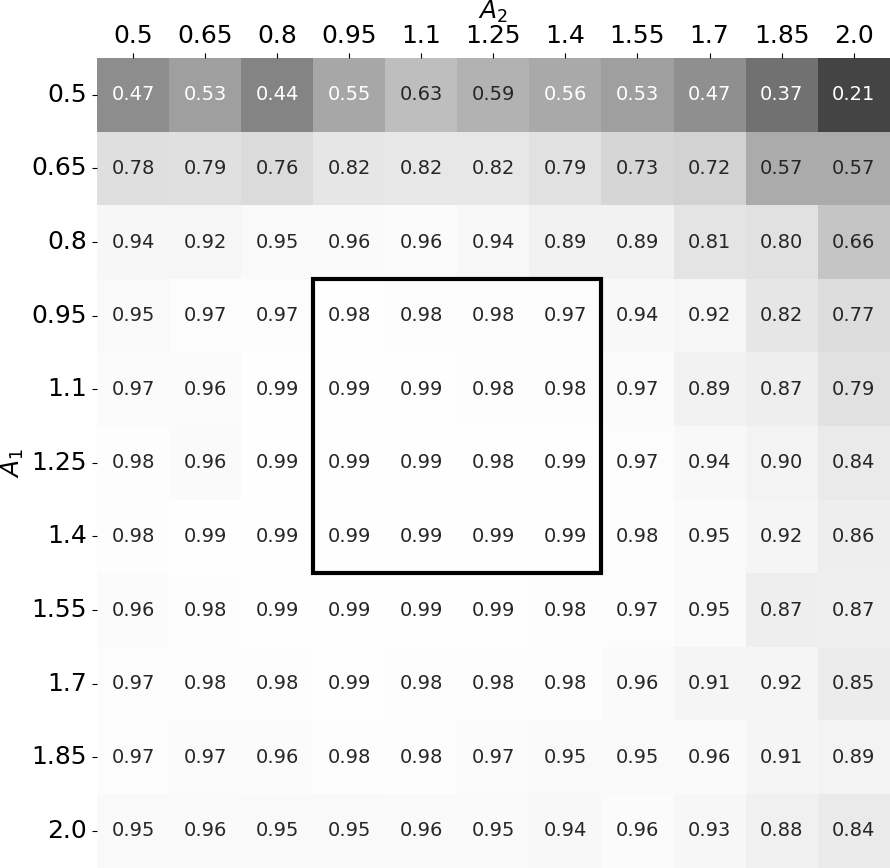} \\ (b)}
    \end{minipage}

    \caption{ 
    The dependence of the coefficient of determination $R^2$ of (a) all parameters, (b) $p_1$ parameter on the relative amplitudes $A_{1,2}$. The coefficient was averaged over 10 runs with random initialization of neural network weights. The black square represents the training data subset. The coefficient of determination is shown by gradation of grey: the lower the coefficient, the darker the color.}
    \label{fig:amplitudes}
\end{figure}

In the second experiment, the ML-model was trained on a newly generated dataset with 50000 samples with the values of the relative amplitudes $A_{1,2} \in [0.95, 1.1, 1.25, 1.4]$. The ML-model was tested on a dataset with 50000 samples from Section 3.3 
with $A_{1,2} \in [0.5, 0.65, 0.8, 0.95, 1.1, 1.25, 1.4, 1.55, 1.7, 1.85, 2]$. All other parameters were randomly changed. The model architecture and training parameters are the same as those used in Section 3.2. 
The results of the experiments are shown in Fig.~\ref{fig:amplitudes}.

It can be seen that with an increase in the difference between the values of the relative amplitudes used during training and testing (when approaching the boundaries of the square in Fig.~\ref{fig:amplitudes} (a)), the accuracy decreases. At the same time, the accuracy in most cases remains acceptable, which indicates the ability of the network to generalize the information received without over-fitting for specific values of latent parameters. 

An error for different parameters $p_i, q_i$ contributes to the overall accuracy differently depending on values of latent parameters $A_{1,2}$. If we analyze in more detail the accuracy of reconstructing the parameters $p_1$, we can see that the accuracy falls sharply if the value of the parameter $A_1$ decreases (see Fig.~\ref{fig:amplitudes} (b)). In this case, the relative contribution of this spectral component to the energy flux becomes smaller, which complicates the determination of its parameters. For parameters $p_2$ and $q_2$ the situation is similar: when the value of the parameter $A_2$ takes on minimum values, the accuracy of reconstructing the parameters decreases. In the case of parameters $p_0$ and $q_0$, the accuracy deteriorates when the values of the parameters $A_1$ or $A_2$ are close to the maximum values. Thus, it can be summarized that in the case of a change in the relative amplitudes of various spectral components, the key factor is the distinguishability of the corresponding spectral component against the background of others. In other words, the relative energy contribution of the spectral component to the cumulative energy flux should be noticeable.

\section{Discussion}
\label{sec:discussion}
This article is devoted not only to solving the problem of reconstructing the tilt parameters of a laser pulse, but to a greater extent to studying the generalization ability of the ML-model. The results of the experiment, when the ML-model is trained on some values of latent parameters, and tested on other values, can be treated differently. From the point of view of latent parameters, the available information is extrapolated by the ML-model beyond the set of parameters values used for training. Generally speaking, the possibility of solving such type of problems with high accuracy is not guaranteed, because the data distribution is different on training and testing datasets, which generally leads to the dataset shift problem. Fortunately, in the problem under consideration, it is the values of the reconstructed tilt parameters that mainly determine the energy flux distribution, while the latent parameters lead only to its modification. This fact explains the good generalization ability of the model, which makes it possible to optimistically assess the possibility of its application in the case when the values of latent parameters are unknown and, as an ultimate case, to real experimental data.

In the context of applying the trained ML-model to real experimental data, it should be noted that in this work a simplified pulse model was used, when the entire spectrum of the laser pulse was divided only into three spectral components. In the case of adapting the model to a real physical experiment, when the wavefront has a complex structure, such a partition must be done into a larger number of spectral components. In this regard, it is important to note that in our numerical experiments it was observed a systematic effect of deterioration in the accuracy of reconstructing the parameters of spectral components with a small wavelength or relative amplitude. In addition, the effect of reducing the accuracy was observed in the case when the boundary wavelengths become close, i.e. one of the spectral ranges becomes very narrow. In all these cases, the influence of such spectral components on the energy flux decreases, which makes it difficult to reconstruct the corresponding tilt parameters. In addition, cross-influence of spectral components with close wavelengths on each other is possible, when the values of tilt parameters from one spectral component can be erroneously attributed by the ML-model to another. The probability of this kind of errors increases sharply with a decrease of the spectral component width, because the relative energy contribution to the cumulative energy flux will also decrease. This means that in the case of partitioning into a large number of spectral components, the characteristics of which may be close, the accuracy of simple ML-models may deteriorate significantly, which will require the use of more complex ML-models or approaches.

\section{Conclusion}
\label{sec:conclusions}
In this paper an ML-based approach for determining the parameters of the spectral-dependent tilt of a focused laser pulse based on the cumulative energy flux in the focal plane was proposed. A simplified formulation of the problem was given with the division of the spectrum of a laser pulse into three spectral components with an individual setting of the tilt parameters for each of them. It was shown that it is possible to obtain good accuracy of tilt parameters reconstructing using the convolutional neural network. In order to study the generalization ability of the proposed ML-model the methodology based on the study of the influence of the latent parameters choice on the accuracy of the ML-model was used. For this purpose separate datasets with different subsets of the latent parameters values were used during training and testing. Two experiments were carried out in which the boundary wavelengths and relative amplitudes acted as such latent parameters. In both experiments, it was shown that the network has good generalization ability and can reconstruct the tilt parameters with quite good accuracy even for the latent parameter values not included in the training set. Significant degradation of accuracy was observed only in cases where the relative energy contribution of the spectral component to the energy flux became small. These results indicate that the proposed approach is promising even in the presence of uncertainty in the choice of the latent parameter values that are typical for application of analytical models to a real experiment. 

\begin{acknowledgments}
This research was funded by the Ministry of Science and Higher Education of the Russian Federation, agreement number 075-15-2020-808. The authors would like to thank Julien~Ferri for useful discussions and other contributions.
The authors acknowledge the use of computational resources provided by the Lobachevsky University and Joint Supercomputer Center of the Russian Academy of Sciences.
\end{acknowledgments}

%
%

\end{document}